\documentclass[]{spie}  %>>> use for US letter paper
\usepackage[]{graphicx}
\usepackage{hyperref}

\title{Telescope Bibliographies: an Essential Component of Archival Data Management and Operations}

\author{Alberto Accomazzi,  Edwin Henneken, 
Christopher Erdmann, and Arnold Rots
\skiplinehalf
Smithsonian Astrophysical Observatory, 60 Garden Street, Cambridge, MA, 02138 USA
}

\authorinfo{Further author information: (Send correspondence to A.A.)\\  A.A.: E-mail: aaccomazzi@cfa.harvard.edu\\  E.H.: E-mail: ehenneken@cfa.harvard.edu\\  C.E.: E-mail: cerdmann@cfa.harvard.edu\\  A.R.: Email: arots@cfa.harvard.edu}

\begin{document} 
  \maketitle 

%%%%%%%%%%%%%%%%%%%%%%%%%%%%%%%%%%%%%%%%%%%%%%%%%%%%%%%%%%%%% 
\begin{abstract}
Assessing the impact of astronomical facilities rests upon an
evaluation of the scientific discoveries which their data have
enabled. Telescope bibliographies, which link data products with the
literature, provide a way to use bibliometrics as an impact measure
for the underlying observations. In this paper we argue that the
creation and maintenance of telescope bibliographies should be
considered an integral part of an observatory's operations. We review
the existing tools, services, and workflows which support these
curation activities, giving an estimate of the effort and expertise
required to maintain an archive-based telescope bibliography.

\end{abstract}

%>>>> Include a list of keywords after the abstract 


%%%%%%%%%%%%%%%%%%%%%%%%%%%%%%%%%%%%%%%%%%%%%%%%%%%%%%%%%%%%%
\section{INTRODUCTION}
\label{sec:intro}  % \label{} allows reference to this section

A well-established way to assess the scientific impact of an
observational facility is the quantitative analysis of the studies
published in the literature which have made use of the data taken by
the facility.  A requirement of such analysis is the creation of
telescope bibliographies which annotate and link data products with
the literature, thus providing a way to use bibliometrics as an impact
measure for the underlying data. Creating such links and
bibliographies is a complex task\cite{2010ASPC..433..109K}.
It is a laborious process which involves specialists searching the
literature for names, acronyms and identifiers, and then determining
how observations were used in those publications, if at all. Depending
on the circumstances, these specialists may be either librarians or
archivists.  For the sake of simplicity, we will refer to them as
bibliography curators in the remainder of this paper. 

The creation of such links represents more than just a useful way to
generate metrics: doing science with archival data depends on being
able to critically review prior studies and then locate the data used
therein, a basic tenet behind the principle of scientific
reproducibility.  From the perspective of a research scientist, the
data-literature connections provide a critical path to data discovery
and access.  Thus, by leveraging the efforts of librarians and
archivists, we can make use of telescope bibliographies to support the
scientific inquiry process.

The SAO/NASA Astrophysics Data System (ADS) provides services which
support both activities described above: bibliography curators use the
ADS's search capabilities to discover and tag data products appearing
in the literature, and the ADS harvests the data-paper links that have
been generated by them to enable the discovery of data products 
from bibliographic records.  Additionally, by integrating these bibliographies
in its search capabilities, the ADS abstract service allows one to filter
literature searches based on bibliographic groups, allowing for queries
such as ``give me all the papers about blazars that have HST and Chandra data.''

In this paper we review the existing and planned
tools, services, and workflows which support the activities of
a telescope bibliography curator.  
We then provide guidance on the effort and skills
required to maintain a bibliography and conclude with a series
of recommendations for projects considering these activities.

\section{ADS Services}
\label{sec:ads}

The ADS databases\footnote{\url{http://adsabs.harvard.edu/ads\_abstracts.html}} 
provide a search interface over bibliographic data covering the entire
astronomical literature (1.9 million records), as well as a large
portion of the physics and optics literature.  As of this writing, the
ADS database contains a total of over 9 million bibliographic records.
These records are linked to a variety of resources, including
astronomical objects, article-level citation metrics, and data
products from a select number of archives.  The ADS is being used on a
daily basis by research scientists and librarians alike.  Because of
its completeness, it is considered the reference site for scholarly
literature in astronomy and is being used by a growing number of
astronomical institutions and projects all over the world to create
and maintain bibliographies.

\begin{figure}
   \begin{center}
   \begin{tabular}{c}
   \includegraphics[width=5in]{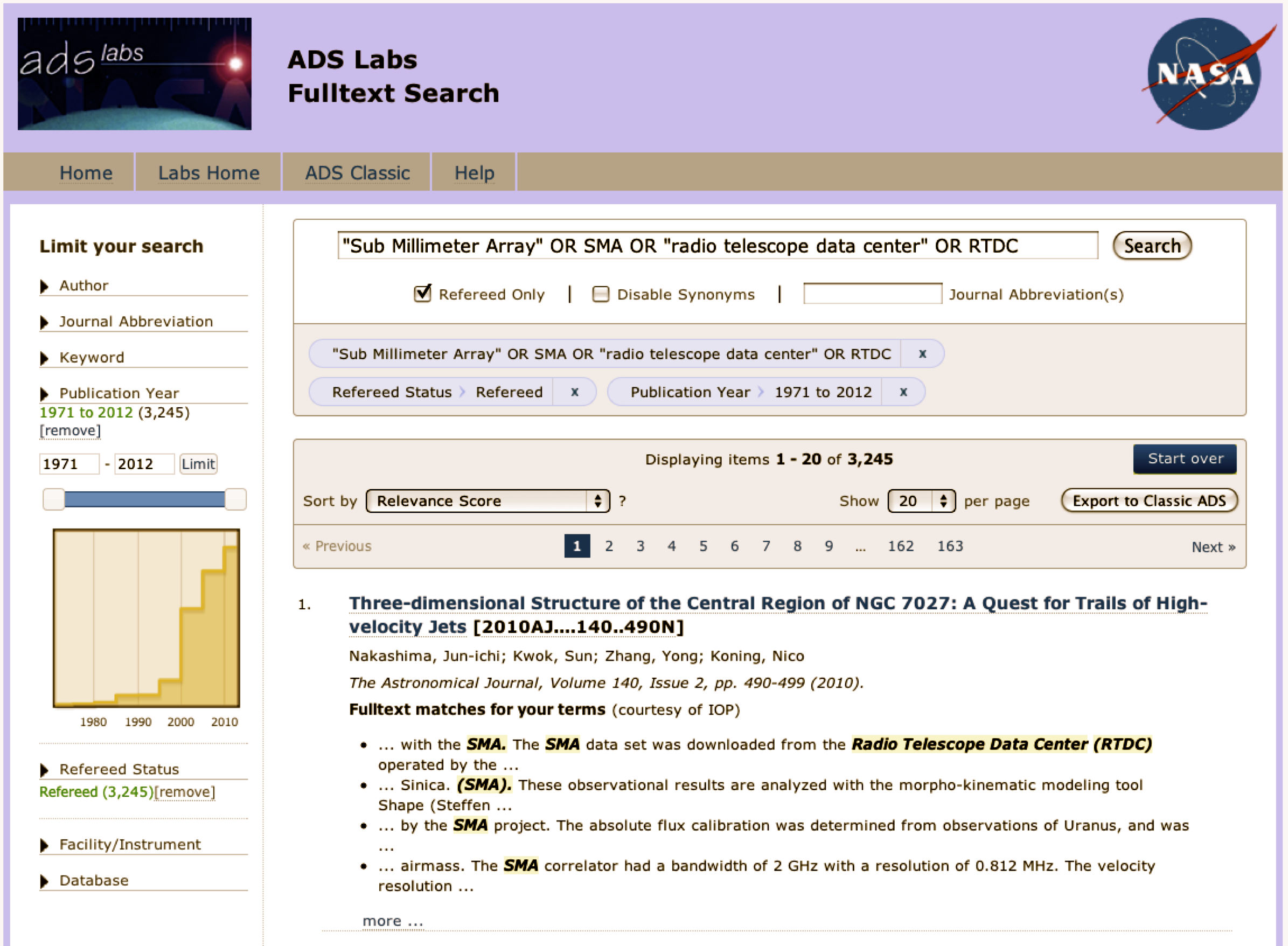}
   \end{tabular}
   \end{center}
   \caption[Fulltext] 
   { \label{fig:fulltext} 
The ADS full-text search allows one to search for specific terms in the
body of the full-text literature archive maintained by ADS. 
The service returns, for each of the matching papers, a list of 
``snippets'' of text highlighting the context in which the search terms were found.
A list of facets in the left column of the screen allows the user to further restrict
the results by selecting additional bibliographic filters.  In this case, we have
constrained the results by restricting publication date range and requesting 
refereed papers only.
   }
   \end{figure}

The ADS also maintains a full-text archive of current and historical
publications covering the main astronomical literature (approximately
1 million documents) 
and several physics journals, which account for another million documents.  
Thanks to agreements between the ADS and the major publishers of astronomy and physics 
literature, this full-text archive now includes current content as it is being published, 
and this content is being indexed and made searchable through a single, unified interface, 
accessible from the ADSLabs full-text service\footnote{\url{http://adslabs.org/fulltext}}.  
The search interface (shown in Figure~\ref{fig:fulltext}) allows users
to find all instances of particular words or phrases in the body of
the articles indexed and returns, for each of the matching papers, a
list of “snippets” of text highlighting the context in which the
search terms were found.  The project is planning to develop an API
that can be used by collaborators to perform text-mining activities
which will facilitate the creation and maintenance of bibliographic
databases.

An analysis of the ADS search logs shows that the greatest use of the
system comes from searches focusing on papers written by a particular
author or on a particular topic.  This search mode typically works
well for researchers since the list of contributors and concepts
covered in a study are summarized in the basic article metadata:
authors, title, keywords, and abstract.   However, the ability to
search the full-text of these publications allows scientists and
bibliographic curators to dig deeper and identify papers which mention
particular entity names such as facilities, instruments, observations,
proposals, or grant numbers.  This provides an effective (as well as
essential) way to generate a rich set of papers
which mention facilities, instruments and data products generated from 
one or more projects.  While the
availability of a single full-text search interface greatly
facilitates the process of finding potentially relevant papers,
it is only the beginning of the bibliography curation process,
which requires additional steps described in the following section.

\section{LINKING DATA AND LITERATURE}
\label{sec:links}

Linking observational metadata with the scientific literature begins
with gathering a rich set of keywords and stop words. Curators gather
as many terms as they can that describe a given project with the
assistance of project scientists. In the case of the SAO
Sub-Millimeter Array (SMA) project, the curators would compile a list
of keywords ranging from ``Pu'u Poli'ahu'' on ``Mauna Kea'' in ``Hawaii'' to
the ``Radio Telescope Data Center'' archive or even ``SMA correlator.''
Typically, curators cast their net out wide and use the
exhaustive list of keywords to scan the full-text
of the journal literature for any mentions of authors, affiliation,
facilities, data, techniques or anything else particular to the
project of interest. Stop words can be used contextually to counteract
and filter the false hits that would accompany the broader keyword
search.  As an example, since SMA is an acronym used in physics to
refer to Shape Memory Alloys, a curator may use the term ``alloy'' as a
stop word to exclude a text snippet containing both ``SMA'' and ``alloy.''

Armed with an exhaustive list of keywords and stop words, the curator
would use a semi-automated search and discovery tool which typically
relies on the ADS to discover papers containing relevant content.  A
few such systems exist today:  FUSE, developed by the European
Southern Observatory\cite{2010ASPC..433...81E}; NRAOPapers,
developed by NRAO and currently being used by its librarians; and the
Chandra bibliographic archive\cite{2010ASPC..434..461W}. In a
systematic fashion, the curator would create subsets of journal
metadata, typically for a date range or by certain journal titles,
using the ADS search interface. Metadata from the ADS, and occasionally the
publishers, is ingested into the system, searched against the compiled
list of keywords and stopwords, and filtered results are outputted to
a triage zone.

Over time, the curator can advance to a point where he/she can spot an
article in the triage zone that for instance mentions the SMA but only
in reference. It is in this area that the curator weeds out or keeps
articles based on careful deduction, assistance of a project scientist
or ultimately by contacting the authors for further clarification. The
standards for citing observational metadata are largely loose and
authors often provide vague clues that the curator must catch to
accurately tag the work. For example, authors could make the claim that they used
submillimeter array data in their study but could be referencing a
similar project such as ALMA.

Once the curator has determined that an article belongs in the project
collection or bibliography, a record is formed from metadata pulled
from the ADS and associated metadata is tied to the record by creating
additional annotations (``tagging''). This activity allows a curator to
enrich the basic bibliographic metadata with items of interest to his
or her project.  This may include the proposal cycle under which
observational data was obtained, number of antennas used in the study,
frequencies used, etc. The driving force behind this
activity is to learn from current science programs such as the SMA and
develop better programs for future science activities. Ultimately, the
projects use the metrics that result from the latter curatorial
activities to review, assess and benchmark projects.

\section{A VALUE-ADDED CURATION LIFECYCLE}
\label{sec:curation}

While the main goal of an archivist or librarian who engages in the
creation of a bibliography is to provide a rich set of relevant
statistics to management, the information collected can be repurposed
for a variety of uses that go beyond its original intended purpose.
For instance, a facility director might be primarily interested in
generating metrics to include in annual reports to funding agencies,
whereas a peer-review panel or time allocation committee will look at
the track record of individual observers before awarding additional
telescope time.  Archive scientists will find the bibliography useful
in understanding how data are first published and whether they are
re-used, while research scientists will benefit from discovering
connections between papers and data as well as identifying
multi-wavelength studies.

In support of discovery, the ADS has long provided, as part of its
search interface, a set of filters which allow users to restrict
results by a limited number of bibliographic groups maintained by the
respective set of curators.  This is done by having ADS periodically
retrieve the list of identifiers from the online bibliographies
maintained by the curators from collaborating institutions and
projects.  For example, the Chandra bibliographic group is updated on
a regular basis by the Chandra archive group, and the list of papers
appearing in it is pulled from the Chandra archive every week-end.  A
similar workflow applies to the papers belonging to all the
bibliographic groups supported in ADS, which are listed in
Table~\ref{tab:groups}. 

\begin{table}[h]
\caption{The ADS Astronomy and Astrophysics Abstract Service
  provides a bibliographic group filter, which gives the user the ability
  to have the results of a query limited to a set of papers which
  belong to one of the bibliographic groups defined below. 
  The level of curation, update frequency, and completeness of such
  groups varies widely, and ADS makes no statement about their overall
  accuracy.  The groups that provide an enhanced level of curation
  by providing links to data product are identified in the description column.}
\label{tab:groups}
\begin{center}       
%\begin{tabular}{|l|p{5.4in}|} %% this creates two columns
\begin{tabular}{lp{5.4in}} %% this creates two columns
\hline
\rule[-1ex]{0pt}{3.5ex} 
 {\bf Group Name} & {\bf Group Description and Policies} \\

\hline
%\rule[-1ex]{0pt}{3.5ex} 
~ & ~ \\

ARI & Papers written by researchers at the
Astronomisches Rechen-Institut in Heidelberg. \\

%\hline

%\rule[-1ex]{0pt}{3.5ex}
CfA & Papers written by researchers at the Center
for Astrophysics (CfA) in Cambridge, MA. \\

%\hline

%\rule[-1ex]{0pt}{3.5ex} 
CFHT & Papers written by researchers at the
Canada-France-Hawaii Telescope (CFHT), HI. \\

%\hline

%\rule[-1ex]{0pt}{3.5ex} 
Chandra & Articles that are related to the Chandra
X-ray Observatory, including articles on Chandra-related theory,
software, calibration, instruments, etc. Those articles that present
specific Chandra observations are linked to the data in the Chandra
Data Archive on which they are based. The database is maintained by
the Chandra Data Archive. \\

%\hline

%\rule[-1ex]{0pt}{3.5ex} 
ESO/Lib & This group contains approximately 10,000 books in
the ESO Astronomy Library. \\

%\hline

%\rule[-1ex]{0pt}{3.5ex} 
ESO/Telescopes & Refereed
articles using ESO data. All papers using data from Paranal (VLT,
VLTI, VISTA, VST) or Chajnantor facilities (ESO observing time of
APEX or ALMA) as well as La Silla papers since publication year 2000
reference the program IDs from which the data originated. They are
linked to the corresponding data products located in the
ESO Science Archive. The database is maintained by the ESO Library. \\

%\hline

%\rule[-1ex]{0pt}{3.5ex} 
Gemini & Articles concerning observations made at
the Gemini Observatory \\

%\hline

%\rule[-1ex]{0pt}{3.5ex} 
Helioseismology & Articles about Helioseismology
provided by the Global Oscillation Network Group (GONG). \\

%\hline

%\rule[-1ex]{0pt}{3.5ex} 
HST & Refereed articles using Hubble Space
Telescope data, generated by the Multimission Archive at STScI.
Note: matching HST proposal IDs to published papers is an on-going
project and never completely up to date. All articles are linked to
the data on which they are based. \\

%\hline

%\rule[-1ex]{0pt}{3.5ex} 
ISO & Articles pertaining to the Infrared Space
Observatory. \\

%\hline

%\rule[-1ex]{0pt}{3.5ex} 
IUE & Refereed articles using International
Ultraviolet Explorer data, generated by the Multimission Archive at
STScI. Note: matching IUE proposal IDs to published papers is
an on-going project and never completely up to date. About 70\% of the
articles are linked to the data on which they are based. \\

%\hline

%\rule[-1ex]{0pt}{3.5ex} 
Keck & Papers written about observations made
at the W.M. Keck Observatory. \\

%\hline

%\rule[-1ex]{0pt}{3.5ex} 
Leiden & Papers written by researchers at the
Leiden Observatory. \\

%\hline

%\rule[-1ex]{0pt}{3.5ex} 
LPI & Articles from the complete Lunar and Planetary Institute
(LPI) bibliography, from approximately 1975 to 1994. These are
typically papers about planetary astronomy. \\

%\hline

%\rule[-1ex]{0pt}{3.5ex} 
NCSA/ADIL & Articles contained in the Astronomy
Digital Image Library (ADIL) at the National Center for Supercomputing
Applications (NCSA). \\

%\hline

%\rule[-1ex]{0pt}{3.5ex} 
NOAO & Papers using National Optical Astronomy
Observatory (KPNO and/or CTIO) observations, maintained
by the NOAO librarian. \\
%\hline

%\rule[-1ex]{0pt}{3.5ex} 
NRAO & Articles related to observations from the
National Radio Astronomy Observatory, maintained by the NRAO librarian. \\

%\hline

%\rule[-1ex]{0pt}{3.5ex} 
PhysEd & Articles from journals related
to education in physics and engineering. \\

%\hline

%\rule[-1ex]{0pt}{3.5ex} 
ROSAT & Articles about ROSAT experiments
and/or ROSAT data. \\

%\hline

%\rule[-1ex]{0pt}{3.5ex} 
SDO & Papers related to the Solar Dynamics
Observatory project. \\

%\hline

%\rule[-1ex]{0pt}{3.5ex} 
Spitzer & Articles about the Spitzer Space
Telescope data, instruments, and articles that provide predictions or
models of Spitzer results. \\

%\hline

%\rule[-1ex]{0pt}{3.5ex} 
Subaru & Papers about data taken with the Subaru
Telescope. \\

%\hline

%\rule[-1ex]{0pt}{3.5ex} 
SMA & Papers about the Smithsonian Astrophysical
Observatory's Submillimiter Array. \\

%\hline

%\rule[-1ex]{0pt}{3.5ex} 
USNO & Papers written by researchers at the United
States Naval Observatory. \\

%\hline

%\rule[-1ex]{0pt}{3.5ex} 
VSGC & Papers about variable stars in globular
clusters. \\

%\hline

XMM & Papers about XMM-Newton Telescope
observations.  \\

\smallskip
\rule[-1ex]{0pt}{3.5ex} 
%\hline 
\end{tabular}
\end{center}
\end{table}

\begin{figure}
   \begin{center}
   \begin{tabular}{c}
   \includegraphics[width=5in]{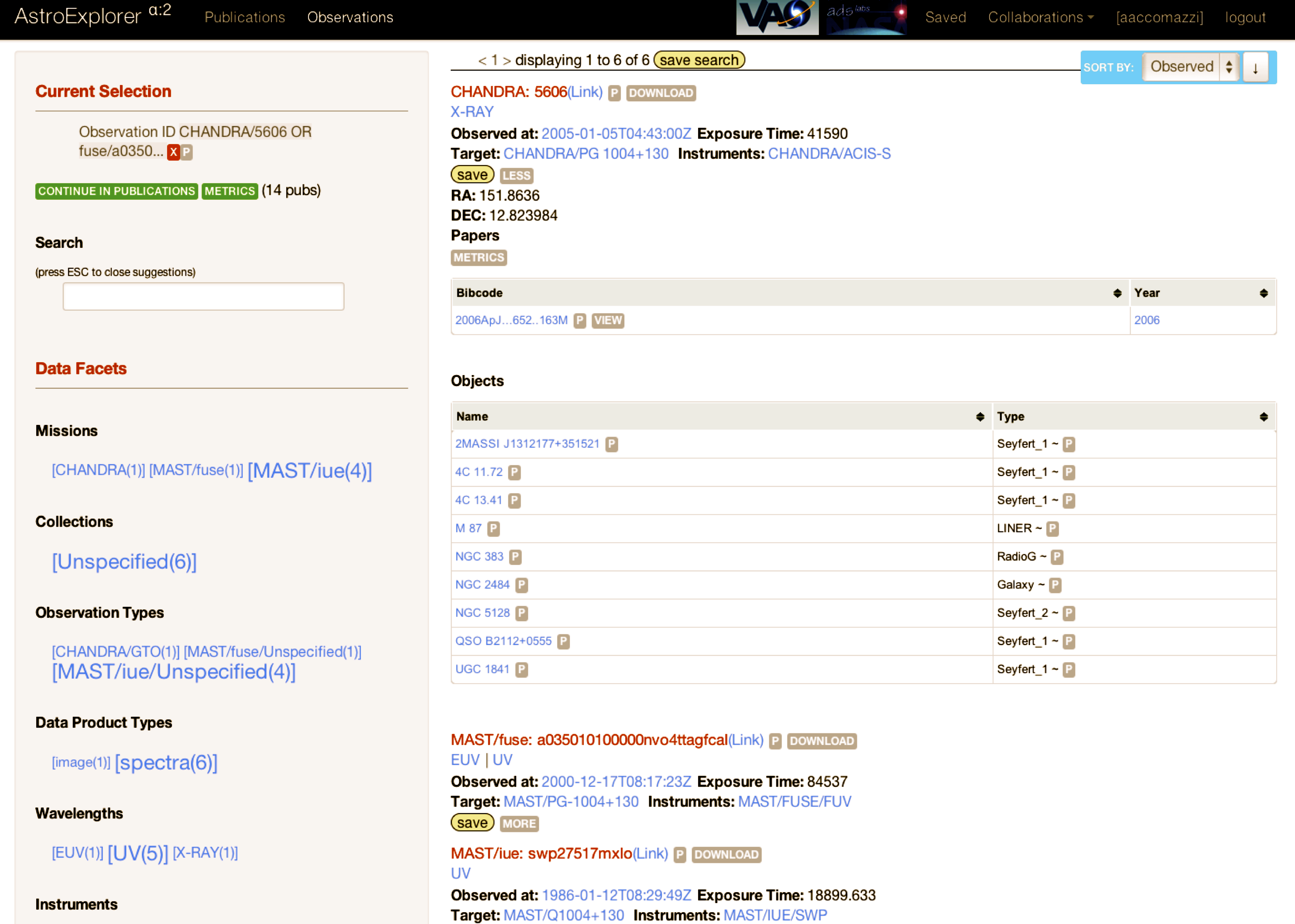}
   \end{tabular}
   \end{center}
   \caption[AstroExplorer] 
   { \label{fig:astroexplorer} 
AstroExplorer is a prototype application which connects observations, objects, and 
the literature.
This screenshot shows a list of Chandra observations (right pane), 
the publication where the first observation was cited and the
objects mentioned in that publication.  The left pane displays data
facets, which can be used to further filter the data products.
   }
   \end{figure}

Observational metadata, astronomical objects and related publications
are things that go well together, and interesting applications can be
built when the connections between them are harvested, modeled, and
indexed.  AstroExplorer\footnote{\url{http://adslabs.org/semantic}} 
is a prototype
application developed by ADS and the VAO 
that helps explore, search, bookmark, share, and work with
assets of importance to astronomers\cite{2011ASPC..442..415A}.  
It harvests and indexes metadata
about, and connections between, papers, data, and astronomical
objects.   The application (shown in Figure~\ref{fig:astroexplorer})
is built upon a semantic database which incorporates linked resources
curated by a number of different archives, specifically: publications
in the ADS; astronomical objects at Centre de Données astronomiques de
Strasbourg (CDS);
and observational metadata from a selected set of telescope
bibliographies (currently Chandra and several MAST missions).  This
type of application is made possible by a close collaboration between
curators at the participating data centers and the AstroExplorer
developers, and currently requires an enhanced level of data sharing
between them.

\begin{figure}
   \begin{center}
   \begin{tabular}{c}
   \includegraphics[width=5in]{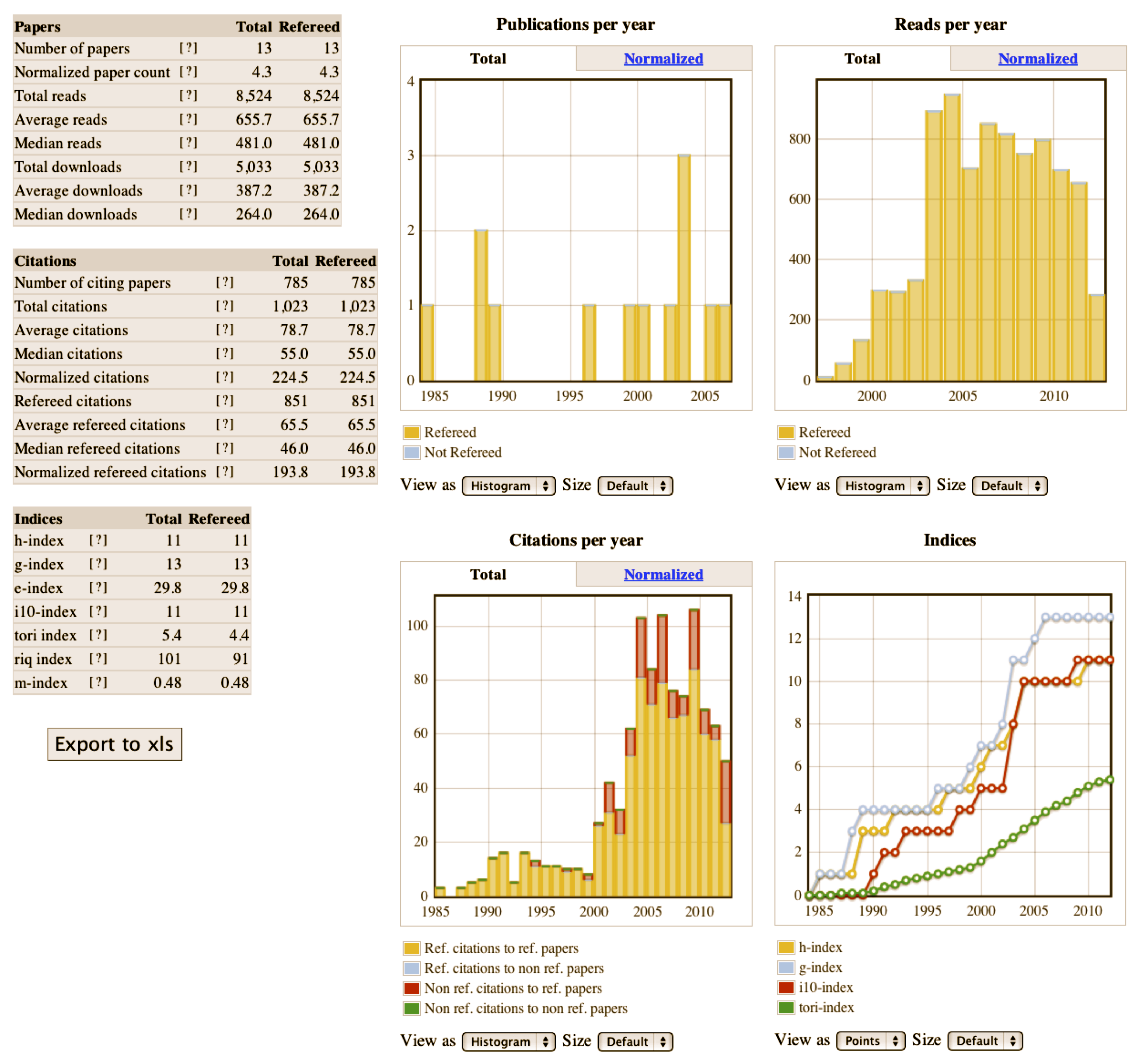}
   \end{tabular}
   \end{center}
   \caption[Metrics] 
   { \label{fig:metrics} 
AstroExplorer uses the ADS metrics API to compute metrics 
related to single observations or a set of observations based on the 
list of papers linked to them.  The metrics currently supported by
ADS fall in three categories: papers published, citations and 
downloads (usage).  For each categories ADS provides various averages,
normalizations and breakdowns in refereed vs. non-refereed.
A set of popular indices and their change over time is also provided.
   }
   \end{figure}

By participating in this joint curation effort and linking data
products to records in the ADS, bibliography curators can take
advantage of the curated metadata and metrics that the ADS provides for all records in its
databases.  Because metrics related to publications are
well-understood in our community, the creation of connections between
data products, observing proposals, grants, and their corresponding
publications provides a straightforward way for a bibliography curator
to use publications as a proxy to evaluate their impact in different
ways.  The typical metrics used for evaluation purposes are the number
of papers published, the number of citations accumulated, and the
usage (downloads) of such papers.  The ADS is able to provide
up-to-date detailed statistics for any of these.
Figure~\ref{fig:metrics} shows a screenshot of the 
metrics summary as generated by AstroExplorer
for the observations shown in Figure~\ref{fig:astroexplorer}.  
The page provides a one-click
download of the metrics data in Excel format and the underlying
machine-readable data is easily accessible from ADS in JSON format through an
API.

The use-cases above provide some examples of the synergy that we gain
from working in a collaborative fashion.  The results of a curation
effort shared between the ADS, the Astrophysics archives and
institutional libraries are clearly greater than the sum of their
parts and are at the heart of the curation lifecycle that our
community enjoys\cite{2010ASPC..433..273A}.

\section{BEST PRACTICES}
\label{sec:practices}

While the post-publication curation efforts outlined above are
essential to maintaining a complete bibliography, much effort can be
saved by moving some of this curation activities up-front, during the
publication process.  There are very good reasons to do so: authors of
a publication know best which data set was used for that publication,
and what software (including versioning).  The principle of research
reproducibility prescribes that the reader should be able to reproduce
results, given a precise description of the data set(s) and
software used.  Ideally, all this information should be included in the
publication metadata during the manuscript preparation process. In
order for this to happen consistently and habitually, a practice of
data citation will have to be introduced that will be as simple as
citing a paper. 

The archive where the data are hosted would be responsible for making
data citable in the first place and for providing authors with instructions
on how to incorporate data citations in their publications. A good example of
an archive that provides detailed guidelines on how to create citable
data sets, together with policies on how to cite these data sets is
the Planetary Data System\footnote{\url{http://pds.nasa.gov}} (PDS).
Another example is the system adopted by the Astronomy Data Centers
Executive Committee (ADEC) and the 
AAS journals\cite{2004AAS...204.7502E,2011ApSSP...1..135A}.
%Alliance (IVOA, see e.g. Eichhorn et al 2007 and Accomazzi 2011).
Examples of repositories where software archived are the 
Astrophysics Source Code Library\footnote{\url{http://ascl.net}} and the
Astrophysics Software Database\footnote{\url{http://www1.astrophysik.uni-kiel.de/asd/}}.

Is having a clear recipe for data citation enough? Will authors adopt
this practice on a voluntary basis and will it become as habitual as
writing an abstract? Our experience is that this is not the
case. Accomazzi\cite{2011ApSSP...1..135A} concludes that ``unfortunately the adoption and use
of dataset identifiers in the [astrophysics] literature has not been a success
story,'' and the situation in the planetary sciences does not seem to
be much different. Realistically, data citation will become common
practice only after we develop a set of editorial and curatorial tools
to go along with the policies and standards which support this effort.
The publishers will have to promote the use of tools to facilitate the
authoring process 
and a form of citation verification during the editing process.
Archives and projects will have to provide user-friendly options to
support identification, self-archiving, persistency, and branding.  As with all things
that need a level of effort, every party involved will want to know
what's in it for them. Everybody involved will agree that
furthering science is a lofty goal, but this may not provide enough
motivation for people to perform additional chores during the publication process. Luckily
there is an effect that can be easily translated into the currency of
the publication process: citations. It has been shown\cite{2011arXiv1111.3618H}
that publications which are linked to the
associated data products result in higher citation rates. For authors
this means that if they facilitate linking their paper to data and
software, this will likely result in more citations. For publishers,
higher citation rates get translated immediately into higher impact
factors. Links and citations enhance a paper’s discoverability, which
is beneficial for every party involved. 

The other side of the equation is discoverability. Citing data and
software is one thing, being able to find those citations and then
being able to follow links to the cited products is another. Knowing
that a search for such citations will be exhaustive is essential for
the process of compiling instrument bibliographies. Although powerful,
full-text search alone is not sufficiently reliable for tracking the
use of identifiers (such as data set identifiers and grant
numbers). For example, using the full-text search on the AGU website
to find citations of PDS data set identifiers results in many false
positives, which you only find out after opening the full article
text, a time intensive process in itself. The full-text search
provided in ADS Labs provides a better solution in the sense that
search results are shown in context through snippets of text and
regular expressions can be used to select and further filter
results. However, as efficient and clever as search can be, it is no
replacement for old-fashioned curation.

\section{EFFORT REQUIRED}
\label{sec:effort}

The curation of links between the literature and items of value to
management and the community is by far the most time consuming aspect
of maintaining a telescope bibliography. However, before the curation
even takes place, the proper IT infrastructure needs to be in place
together with the necessary buy-in from management, to ensure that the
bibliography receives suitable priority. The bibliography curators
must have access to a System \& Database Administrator to assist with
high level technical support. Once a proper infrastructure has been
set-up, a minimal percentage of time is required from the
Administrator over the course of a year and assistance is requested on
an infrequent basis. This position is typically already part of the
organization’s or project's overall IT support, but an appropriate
amount of  time should be budgeted and dedicated specifically to this
task.

A software developer is required in order to install, customize and
enhance applications supporting the curation efforts, as well as to
facilitate the dissemination and exchange of the resulting
work. Preferably, the system that is ultimately developed is
web-based, so that it can both ingest and exchange metadata with other
systems, such as the ADS and VO archives, using web-based protocols
like  OAI-PMH and TAP.  Ideally, the developer will be familiar with
the entire web-based technology stack (HTTP, XML, XSLT, Ajax), modern
programming languages (python, Java), and database technologies in use
in astronomy (SQL, ADQL). Development time can be reduced by
collaborating with groups that have already created applications to
manage telescope bibliographies, including ESO’s suite of software
(Telbib and FUSE), STScI's MAST bibliographic database, and the Chandra Data Archive
bibliographic database\cite{Winkelman2012}. The Software Developer time can range from a
fraction of an FTE to a full FTE, with the latter ultimately improving the
quality of the bibliography. Embedding the developer with the
curators, and allowing for interaction with project scientists, is
highly recommended and leads to a higher quality end-product.

As mentioned earlier, curating a telescope bibliography can be time
intensive, depending on  the degree of text mining involved and the
amount of communication needed with scientists to evaluate items for
inclusion. The amount of effort can range from a fraction of an FTE
for smaller projects to possibly 2 FTE for a bigger organization with
multiple projects.  While the overall FTE count may be small, the
actual number of people involved in the curation effort is often larger,
with each person contributing his or her particular skills.
For instance, an organization such as ESO has two people
part-time responsible for its bibliography, with one performing the more advanced curation,
mining the literature, matching items to criteria, tagging, and
the other doing macro level curation through the use of scripts,
advanced database queries, as well as the development and maintenance
of software tools. In the ESO scenario, the FTEs are embedded
in the library, which has elevated the library within the
organization. The model is different for the Chandra Bibliography,
where the 1 FTE curator's title is Archive Specialist instead of
Librarian. The effort is shared among a group of Archive Specialists
who are also tasked with other archive operations duties.  In
addition, part of a position is also heavily involved with the
development of the bibliography system used for curation and
dissemination. 

\section{RECOMMENDATIONS}
\label{sec:recommendations}

It should be pretty clear by now that we consider it essential for a
new project or archival initiative to include in its plans the task to
create and maintain a comprehensive bibliography.  It has been
encouraging for us to see this happen more and more often, as projects
realize the value of linking data and literature.  However, only in
few cases is the curation effort as thorough as it can be: as we have
argued in the previous sections, the amount of time and effort
involved in creating deep links between individual observations,
instruments, observing proposals, and the literature can vary widely
and requires much more than just coming up with a list of papers
“related” to the telescope.  If one of the goals is to enable
discovery and re-use of the data products stored in an archive, then
make sure to budget for a robust curation process.  Our first
recommendation is therefore:
{\bf make sure to include the cost of creating and maintaining a 
comprehensive bibliography in your operations from the start, and 
don't underestimate its potential cost and impact.}

An archive manager should also plan to use the bibliography to provide
meaningful statistics about the use of data products to management and
scientists involved in evaluating system operations.  Our advice to
the curators is to collect as much metadata as possible in the
creation of the bibliography and to be prepared to have to compute
statistics that haven’t been used yet or answer questions that haven't
been thought of yet.  For example, Rots et al\cite{2012PASP..124..391R,Rots2012} 
have recently
proposed using fraction of observing time published and archival usage
as metrics which are less observatory-specific than the typical ones
(e.g. number of publications or citations per data product).  As we
see more and more studies analyze data from different observatories,
it may become reasonable and necessary some day to estimate the impact
of each individual observation as a fraction of the papers' total
impact.  The more detailed the bibliography is, the easier it will be
for the curator to compute (or estimate) these metrics.  Our second
recommendation is: 
{\bf collect as much metadata as you can as part of the curation
  process, because you will need it some day.}

As most of our curators are well aware, there are great benefits in
promoting practices which facilitate the identification in the
research literature of items of interest to them such as scientific
artifacts (object names, data products), technical configurations
(instruments, software), and administrative aspects (supporting
grants, proposal ids).  There have been some attempts to standardize
the names of the relevant entities (e.g. object names abide by a
well-established nomenclature) and to provide ways for authors to
properly identify them during the manuscript creation process.  While
these efforts have seen mixed results, we feel it is everybody's
responsibility to try to improve the situation.  A new archival system
which plans to distribute data products to the public should provide
guidance to its users on how the data products are named and
how they should be referenced in the literature.  Similar requirements
should apply to acknowledging observing programs and grants which
enabled the creation and analysis of the observations.
Our third recommendation is: 
{\bf provide clear instructions and requirements to archive users
  which promote the current best practices for citing data, software,
  funding sources, facilities, etc.}

As discussed earlier in this paper, you will not be the first one to
create a telescope bibliography.  Neither will you be the first one to
be tempted to start from scratch and build on the expertise of your
local programmers, archivists and librarians.  While this is an
entirely reasonable way to get the job done, it has the unfortunate
effect of further splintering the number of existing bibliographic
management systems in use in astronomy today.  As mentioned in section
2, there is a growing number of products in use today for this
purpose, and as more projects participate in this effort, it would be
desirable to have at least one general-purpose system which supports the basic
functionality outlined above: search, save, review, annotate.  With
the introduction of the comprehensive full-text search capabilities in
the ADS, there is no reason why a bibliography system cannot be
developed as a web-based application which sits on top of the ADS
search engine, relies on the ADS user accounts, and which connects to
the observatonal database of each participating data archive through
a proper driver or plug-in.  Such system does not yet exist but
we hope that it will in the not too-distant future, 
and should be built through a collaborative software
development process.  Our final recommendation is thus: 
{\bf do not reinvent the wheel but rather work with us in building a
  general-purpose, web-based, modular curatorial ecosystem that will
  benefit the entire community.}

\acknowledgments
This work was partly supported by the ADS, which is funded by NASA grant NNX09AB39G.
We are grateful to Uta Grothkopf for her valuable comments and clarifications
about ESO's curation effort.

\bibliography{2012SPIE_Accomazzi_etal} 
\bibliographystyle{spiebib} 

\end{document}